\documentclass[aps,pra,10pt,twocolumn,showpacs,preprintnumbers,amsmath,showkeys,superscriptaddress]{revtex4-1}
\usepackage{filecontents}
\usepackage[utf8]{inputenc}
\usepackage{amsmath}
\usepackage{amsfonts}
\usepackage{amssymb}
\usepackage{graphicx}
\usepackage{lmodern}
\usepackage{xcolor}

\usepackage[colorlinks=true]{hyperref}
\usepackage[capitalize]{cleveref}
\hypersetup{colorlinks,
citecolor=blue, 
linkcolor=blue 
}

\begin{document}
\title{Surface resonance of the (2\(\times\)1) reconstructed lanthanum hexaboride (001)-cleavage plane: a combined STM and DFT study}

\author{P.~Buchsteiner}
\affiliation{IV. Physikalisches Institut, Georg-August Universität Göttingen, 37077 Göttingen, Germany}
\author{F.~Sohn}
\affiliation{Institut für Theoretische Physik, Georg-August-Universität Göttingen, 37077 Göttingen, Germany}
\affiliation{Institut für theoretische Physik, Technische Universität Clausthal, 38678 Clausthal-Zellerfeld, Germany}
\author{J.~G.~Horstmann}
\affiliation{IV. Physikalisches Institut, Georg-August Universität Göttingen, 37077 Göttingen, Germany}
\author{J.~Voigt}
\affiliation{IV. Physikalisches Institut, Georg-August Universität Göttingen, 37077 Göttingen, Germany}
\author{M.~Ciomaga~Hatnean}
\affiliation{Department of Physics, University of Warwick, Coventry CV4 7AL, United Kingdom}
\author{G.~Balakrishnan}
\affiliation{Department of Physics, University of Warwick, Coventry CV4 7AL, United Kingdom}
\author{C.~Ropers}
\affiliation{IV. Physikalisches Institut, Georg-August Universität Göttingen, 37077 Göttingen, Germany}
\author{P.~E.~Blöchl}
\affiliation{Institut für theoretische Physik, Technische Universität Clausthal, 38678 Clausthal-Zellerfeld, Germany}
\affiliation{Institut für Theoretische Physik, Georg-August-Universität Göttingen, 37077 Göttingen, Germany}
\author{M.~Wenderoth}
\email[Corresponding author: ]{martin.wenderoth@uni-goettingen.de}
\affiliation{IV. Physikalisches Institut, Georg-August Universität Göttingen, 37077 Göttingen, Germany}

\date{\today}

\begin{abstract}
We performed a combined study of the (001)-cleavage plane of lanthanum hexaboride (LaB\(_6\)) using scanning tunneling microscopy (STM) and density functional theory (DFT). Experimentally, we found a (2\(\times\)1) reconstructed surface on a local scale.  The reconstruction is only short-range ordered and tends to order perpendicularly to step edges. At larger distances from surface steps, the reconstruction evolves to a labyrinth-like pattern. These findings are supported by low-energy electron diffraction (LEED) experiments. Slab calculations within the framework of DFT shows that the atomic structure consists of parallel lanthanum chains on top of boron octahedra. Scanning tunneling spectroscopy (STS) shows a  prominent spectral feature at \(-0.6\)\,eV. Using DFT, we identify this structure as a surface resonance of the (2\(\times\)1) reconstructed LaB\(_6\) (100)-surface which is dominated by boron dangling bond-states and lanthanum \(d\)-states.
\end{abstract}

\maketitle

\section{Introduction}

The rare earth hexaborides (REB\(_6\)) are a material class with a common, relatively simple crystalline structure, but widely tunable electronic and magnetic properties. 
For example, dense Kondo behavior is found in CeB\(_6\) \cite{Kondo}, PrB\(_6\) and NdB\(_6\) order antiferromagnetically \cite{geballe}, 
SmB\(_6\) is a Kondo insulator \cite{kasuya, souma}, 
EuB\(_6\) is a ferromagnetic semimetal which exhibits colossal magnetoresistance \cite{pohlit}, and 
YbB\(_6\) is proposed to host topologically protected states without a Kondo mechanism \cite{YbB6}. 
The variety of these phenomena can be traced back to the \(4f\)-occupancy increasing from \(4f^0\) for La up to \(4f^{14}\) in Lu as the rare earth elements.

\begin{figure}[b]
\centering
\includegraphics[scale=0.14]{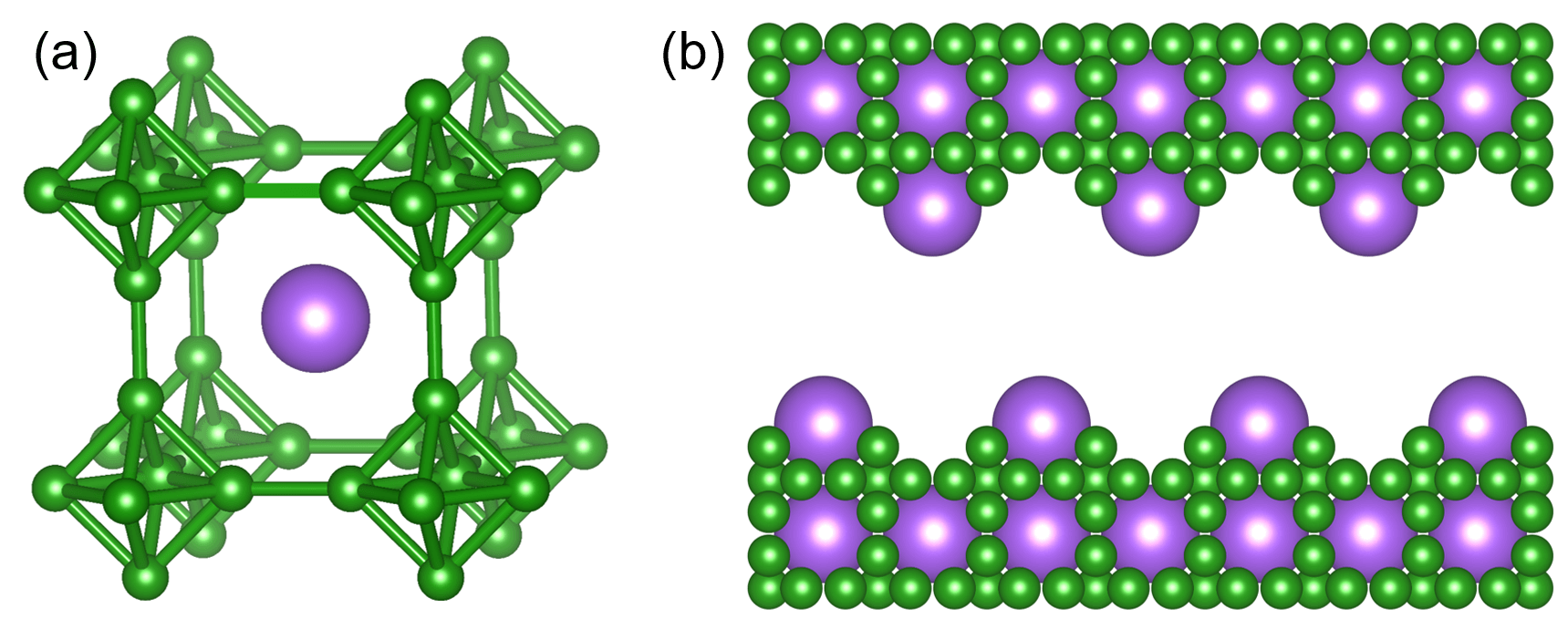}
\caption{(a) Crystal structure of REB\(_6\), where the rare earth element is located at the center of the simple cubic unit cell and the B\(_6\) octahedra at the corners. (b) Side view of two cleaved surfaces, where each side is partly La terminated.}
\label{schema}
\end{figure}

All rare earth hexaborides share the same cubic crystal structure with the B\(_6\) octahedra located at the cube's corners and the rare earth element at the center, as seen in \cref{schema}\,(a). 
A three dimensional covalent binding network between the B\(_6\) octahedra can be achieved by electron donation of the rare earth element, leading to positively charged ions and negatively charged boron cages in the crystal structure \cite{Higgins}. 
The lattice constant changes only slightly across the REB\(_6\) series. 

One of the most prominent hexaborides is LaB\(_6\), a widely used electron emitter due to its extraordinarily low work function \cite{surfacetrenary}. 
Recently there has been a growing interest in its solar heat absorbance with regard to possible applications in solar energy devices \cite{Sani, Mattox}. 
Although surface properties play a crucial role in these applications, surface studies of this system display a rather incomplete picture. 
Even the ground state geometrical structure of the (001)-surface is under current debate. 
Up to the present date, experimental results show a simple (1\(\times\)1) reconstructed surface \cite{surfacetrenary}, as found in low-energy electron diffraction (LEED) and Auger electron spectroscopy (AES) studies \cite{LEED1, LEED2, LEED3, swanson1, swanson2, chambers, goldstein}. 
Scanning tunneling microscopy (STM) experiments at room temperature and under UHV conditions have shown a (1\(\times\)1) structure, which has been described to be lanthanum terminated \cite{ozcomert1, ozcomert2}. 
However, it should be noted that the samples of the aforementioned studies have been prepared by polishing and heating. 
Recently, surface slab calculations of LaB\(_6\) (001) based on density functional theory (DFT) have been made \cite{Schmidt}. Therein, various surface reconstructions are taken into account. 
Their prediction is a (2\(\times\)1) reconstructed surface as ground state. These findings seemingly contradict the (1\(\times\)1) surface structure observed so far. 

To solve this discrepancy between previous experimental results and recent theoretical predictions, we investigated LaB\(_6\) (001) prepared by cleavage in UHV. 
Therefore, annealing-related impacts on the surface morphology, such as preferential evaporation or thermally activated reorganization, can be minimized. 

A (2\(\times\)1) reconstructed surface is unambiguously found by using high-resolution STM and LEED measurements. 
DFT slab calculations resolve the chemical nature of the (2\(\times\)1) reconstruction as parallel rows of lanthanum atoms on top of non-reconstructed B\(_6\) cages. 
Its electronic structure close to the Fermi energy is governed by a surface resonance, which is mainly composed of boron \(sp\)-hybrid dangling bond orbitals.

\section{Methods}
\subsection{Experimental techniques}
STM experiments were performed in a home-built microscope operating at 8\,K and at a base pressure of 4\(\times\)\(10^{-11}\)\,mbar. 
Tunneling tips were made by electrochemical etching of polycrystalline tungsten wire. 
The LaB\(_6\) single crystals were grown using the floating zone technique as described in Refs.\,\onlinecite{Geetha1, Geetha2}.
The crystals were oriented by gamma-ray diffraction and cut along the (001)-plane into rectangular samples of about 1\(\times\)4\,(mm)\(^2\) size and 300\,\(\mu\)m thickness. 
The samples were cleaved in situ at room temperature along the (001)-plane followed by immediate transfer to the STM head at cryogenic temperature. 
All STM images were recorded using the constant current topography (CCT) mode. 
LEED experiments were performed on identically prepared samples, cleaved at a base pressure of 2\(\times\)\(10^{-8}\)\,mbar and investigated at 2\(\times\)\(10^{-10}\)\,mbar. 
The diffraction images presented in this work were recorded at either room temperature or at 27\,K. 
For our LEED experiments we used an ultrafast LEED setup (ULEED), as described in Ref.\,\onlinecite{uleed2}. 
This setup features a laser-pulsed electron gun with an electron beam diameter of about 80\,\(\mu\)m at the sample. 
With this technique, the electron beam contains significantly less electrons than in convential systems. 
Thus, the possibility of electron beam induced surface damage is drastically reduced. 
The resulting small number of scattered electrons is detected with a microchannel plate (MCP). 
Atomic force microscopy (AFM) measurements were conducted in a commercial instrument manufactured by Agilent, which operates at ambient condition. 

\subsection{Calculations}
Theoretical results presented in this work are based on DFT \cite{hohenberg, kohn} and are obtained with the CP-PAW code \cite{cppawhome}, which employs the projector-augmented wave method \cite{blochl} together with a functional minimization scheme derived from the Car-Parrinello molecular dynamics approach \cite{car}. 
We use the local hybrid exchange-correlation functional PBE0r described elsewhere \cite{sotoudeh}, which locally replaces a fraction of the PBE \cite{perdew} exchange with the same portion of the exact Hartree-Fock exchange. In the PBE0r functional the Fock-term is expressed in local orbitals and only on-site terms are retained. 

\section{Results}

\subsection{Surface Morphology}
\begin{figure}[htbp]
\centering
\includegraphics[scale=0.2]{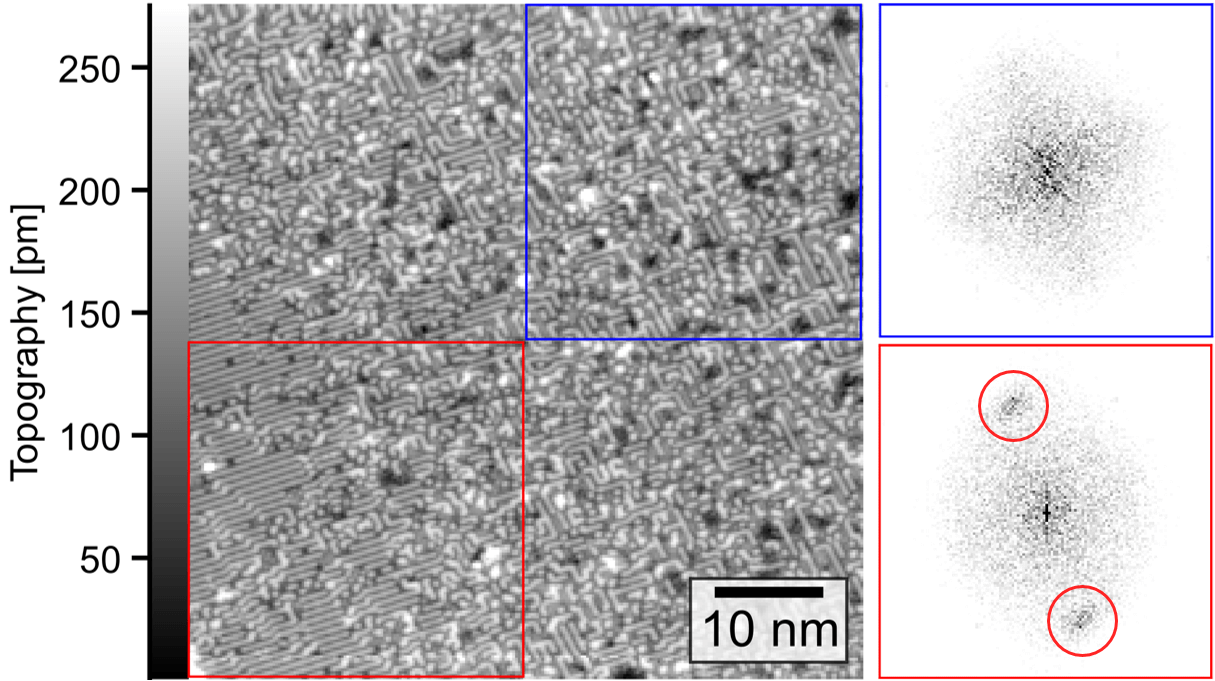}
\caption{Atomically resolved CCT taken at 0.8\,V/0.1\,nA. Atomic rows with a spacing of two bulk lattice constants are present, which are mostly labyrinth-like arranged. The Fourier analysis of a surface area with a rather ordered region shows clear signs of the (2\(\times\)1) reconstruction, see red-marked area. For most regions no distinct peaks in the Fourier transform can be observed, see blue-marked area.}
\label{FFT}
\end{figure}

For most of the encountered surfaces probed by STM, the surface appears rather disordered. 
This is described in more detail in \cref{Supporting_experimental_data}. 
Atomically ordered areas on LaB\(_6\), as seen in \cref{FFT}, are scarce and have to be searched for. 
The atomic structure appears chain-like with a spacing of two bulk lattice constants. 
Hence, (2\(\times\)1) reconstruction peaks can be clearly observed in the Fourier analysis, as seen for the red-marked area of \cref{FFT}. 
However, this (2\(\times\)1) reconstruction is ordered only on a short range and is mainly labyrinth-like arranged, as seen in the upper right corner of the CCT image in \cref{FFT}. 
Although individual chains can still be resolved, no signs of a (2\(\times\)1) reconstruction can by found in the Fourier analysis due to the lack of long-range order. 

\begin{figure}[htbp]
\centering
\includegraphics[scale=0.14]{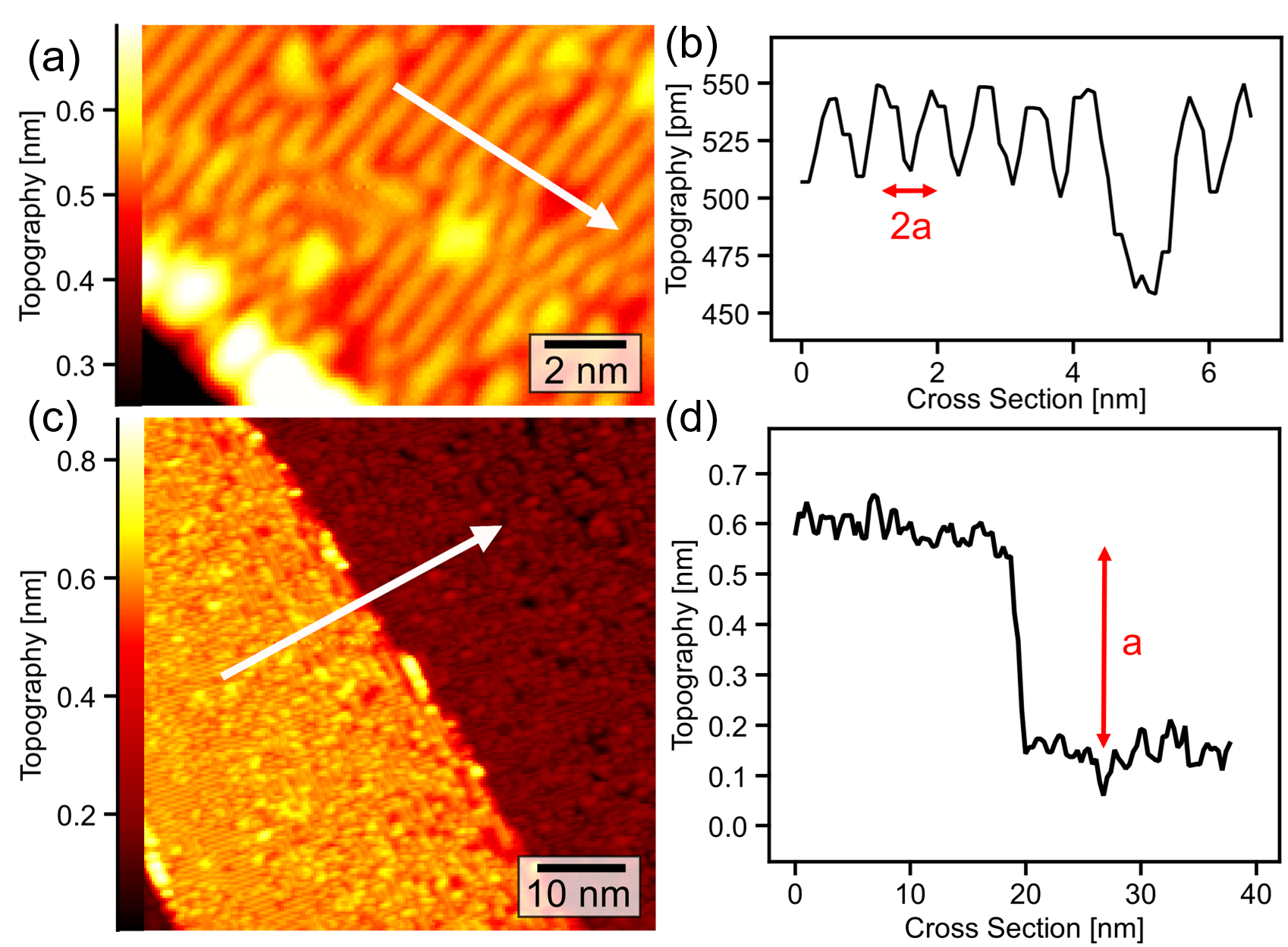}
\caption{(a) High-resolution CCT of the (2\(\times\)1) reconstruction taken at 1\,V/0.1\,nA. (b) Cross section along the (2\(\times\)1) reconstruction as indicated by the white arrow in (a). (c) Large scale overview taken at 1\,V/0.1\,nA in the vicinity of the high-resolution image in (a). (d) The height profile of (c) shows that the step is of about 4.1\,\AA\;height, which amounts to one bulk lattice constant \(a\).
}
\label{stepsA}
\end{figure}

A high-resolution image of the (2\(\times\)1) reconstruction, \cref{stepsA}\,(a), shows that even for the more ordered areas, the chains exhibit kinks and defects. 
In the vicinity of the region shown in \cref{stepsA}\,(a) a step edge of one lattice constant height has been found. 

\begin{figure}[htbp]
\centering
\includegraphics[scale=0.23]{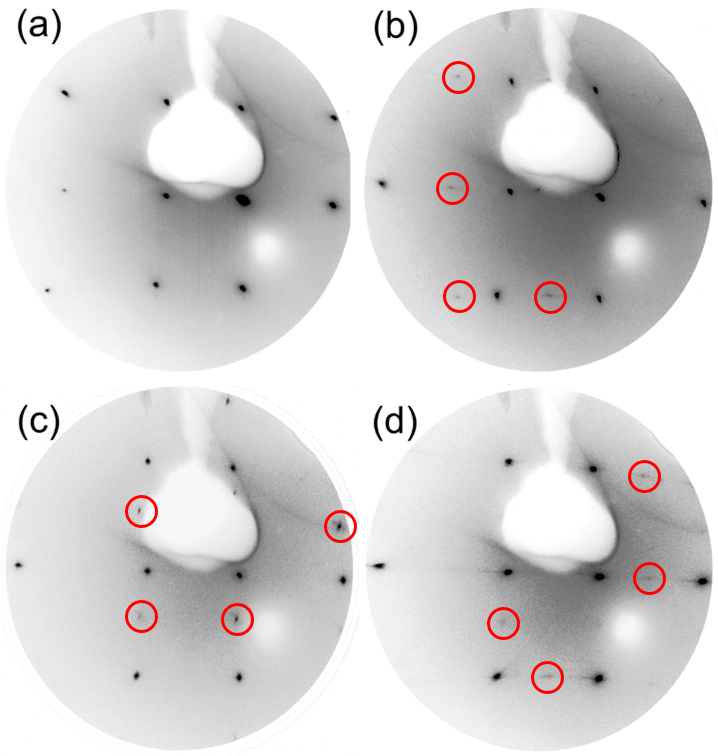}
\caption{ULEED images of different locations on the cleaved LaB\(_6\) (001)-surface. (a) Only (1\(\times\)1) spots are visible, taken at 130\,eV and room temperature. In (b) additionally (2\(\times\)1) spots are present, 100\,eV/RT and in (c) (1\(\times\)2) spots are observed, 100\,eV/27\,K. In (d) both (1\(\times\)2) and (2\(\times\)1) spots are present, 100\,eV/27\,K. The observed reconstruction spots are marked by red circles.}
\label{ULEED}
\end{figure}
At the step edge, the most ordered (2\(\times\)1) reconstruction is observed. 
Farther away from the step, the chains lose their preferential orientation perpendicular to the step edge, and a more labyrinth-like arrangement is seen, which is similar to the observation in \cref{FFT}. 
Therefore, our findings so far suggest a correlation between steps and the occurrence of a well-ordered (2\(\times\)1) reconstruction. 

The spatially rather limited observation of long-range order explains why signs of a (2\(\times\)1) reconstruction have not yet been found in LEED experiments. 
In our standard LEED setup, only a (1\(\times\)1) structure could be seen that vanished after about 30 minutes of measurement time at a pressure of \(10^{-9}\)\,mbar. 
The situation changes when using the ultrafast LEED setup and measurements could be carried out over a couple of hours without any noticeable change. 
\cref{ULEED} sums up the observed diffraction patterns. 
For most of the sample areas only a (1\(\times\)1) pattern were found, as seen in \cref{ULEED}\,(a). 
In some surface areas, as shown in \cref{ULEED}\,(b) and \cref{ULEED}\,(c), signs of a (2\(\times\)1) or (1\(\times\)2) reconstruction are present. 
In one surface region both (2\(\times\)1) and (1\(\times\)2) spots were observed, as depicted in \cref{ULEED}\,(d). 
Note that the diffraction patterns do not change upon cooling the sample down to 27\,K, apart from the increasing spot intensity and reduced background noise level due to the temperature-dependent Debye-Waller factor. 

To clarify the chemical nature of the (2\(\times\)1) reconstructed surface, scanning tunneling spectroscopy (STS) was carried out.

\subsection{Spatially resolved spectroscopy}
\begin{figure}[b]
\centering
\includegraphics[scale=0.125]{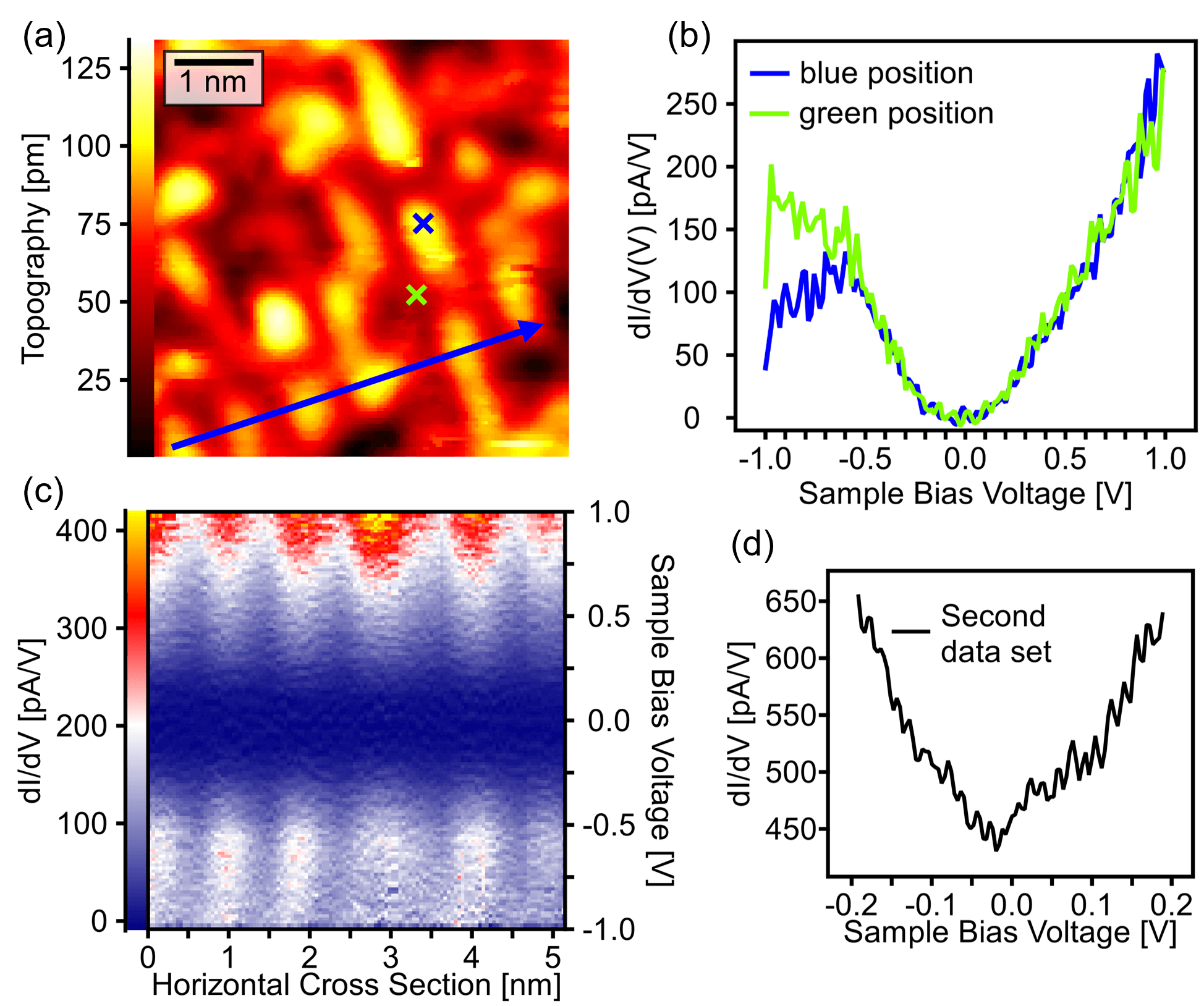}
\caption{(a) CCT taken at 1\,V/0.1\,nA. (b) Two exemplary d\(I/\)d\(V(V)\) curves obtained from the marked positions are shown. (c) The horizontal arrow in (a) marks the line along the d\(I/\)d\(V(V)\) cross section was taken, which is shown here. (d) d\(I/\)d\(V(V)\) curve obtained from a second data set, see \cref{Supporting_experimental_data}, to resolve the energy region around \(E_{\text{F}}\).}
\label{dIdV}
\end{figure}

\cref{dIdV}\,(a) shows a CCT image, where, simultaneously to the topography, at every measurement point an \(I(V)\) curve and the apparent barrier height \(\Phi_{\text{app}}\) were recorded. 
The chains of the (2\(\times\)1) reconstruction in \cref{dIdV} extend for only a few unit cells and are interrupted by various defects. 
Using the \(I(V)\) curve, its differential conductance d\(I/\)d\(V(V)\) can be seen as an approximation for the local density of states (LDOS) \cite{tersoff1, tersoff2}. 
\cref{dIdV}\,(b) shows two d\(I/\)d\(V(V)\) spectra obtained from the blue and the green-marked positions. 
Clearly, a peak in the differential conductance at about \(-0.6\)\,eV can be resolved on top of the protrusion of the reconstruction, blue-marked position, as well as a steep rise of d\(I/\)d\(V(V)\) towards positive bias voltages.
To visualize the spatial variation of the LDOS, the differential conductance can be plotted position dependent as a color coded d\(I/\)d\(V(V,x)\) cross section. 
Since the \(I(V)\) curves are recorded quasi-simultaneously to the constant current topography, these curves are taken on a modulated contour line given by the topography. 
To project the measurement onto a constant height above the surface, a topography normalization is applied, as described in Ref.\,\onlinecite{Garleff}. 
This can be done by using the apparent barrier height \(\Phi_{\text{app}}\), which is defined as \cite{chen}
\begin{align}
\Phi_{\text{app}}=\frac{\hbar^2}{8 m_\text{e}}\left(\dfrac{\text{d}\,\text{ln}I}{\text{d}s}\right)^2 .
\label{apparentbarrierheight}
\end{align}
Here, $\text{d}s$ is the change of the tip-sample separation and $m_\text{e}$ the electron mass. 
The apparent barrier height is often used as an estimate for the sample work function. 
However, its absolute value is connected to the work functions of both tip and sample. 
The spatially resolved \(\Phi_{\text{app}} (x,y)\) map can be seen in \cref{Supporting_experimental_data}, \cref{sts1V}\,(b), with a mean value of (1.05 \(\pm\) 0.17)\,eV. 
After performing the normalization, the d\(I/\)d\(V(V,x)\) data reveals that the \(-0.6\)\,eV peak is strongest at the protrusions of the reconstruction, as seen in \cref{dIdV}\,(c) for the cross section along the marked direction. 
However, for this data set the tunneling current has dropped below 1\,pA in the vicinity of the Fermi energy, which could be mistaken for a non-metallic surface. 
To resolve the energy region around \(E_{\text{F}}\), another spectroscopy was performed, which was acquired at a smaller bias voltage set point of 0.2\,V. 
The spatially averaged d\(I/\)d\(V(V)\)-curve can be seen in \cref{dIdV}\,(d) and the full data set in \cref{Supporting_experimental_data}, \cref{sts200mV}. 
A finite conductance at \(E_{\text{F}}\) is clearly present. 
Moreover, the differential conductance has parabolic shape with a minimum shifted slightly towards negative bias voltages with an additional d\(I/\)d\(V(V)\) feature at 0.1\,eV.
The derived apparent barrier height is \(\Phi_{\text{app}}\) = (2.99 \(\pm\) 0.27)\,eV.

\subsection{Surface simulations}
\label{sec:surface_simulations}

Our experimental findings of a (2\(\times\)1) reconstructed surface together with previous theoretical predictions \cite{Schmidt} point towards a lanthanum terminated (2\(\times\)1) surface reconstruction. Based on density functional theory, we performed an in-depth analysis of the electronic surface structure of such a (001)-surface of LaB\(_6\) with linear chains of lanthanum atoms at the surface, which are separated by void lines. This termination makes the surface formally charge neutral. Details about the DFT simulations, including the unit cell setup as well as technical parameters are given in \cref{sec:appendix_DFT_simulation_details}. Therein, we also present the relaxed surface structure.

\begin{figure}[htb]
\centering
\includegraphics[width=.39\textwidth]{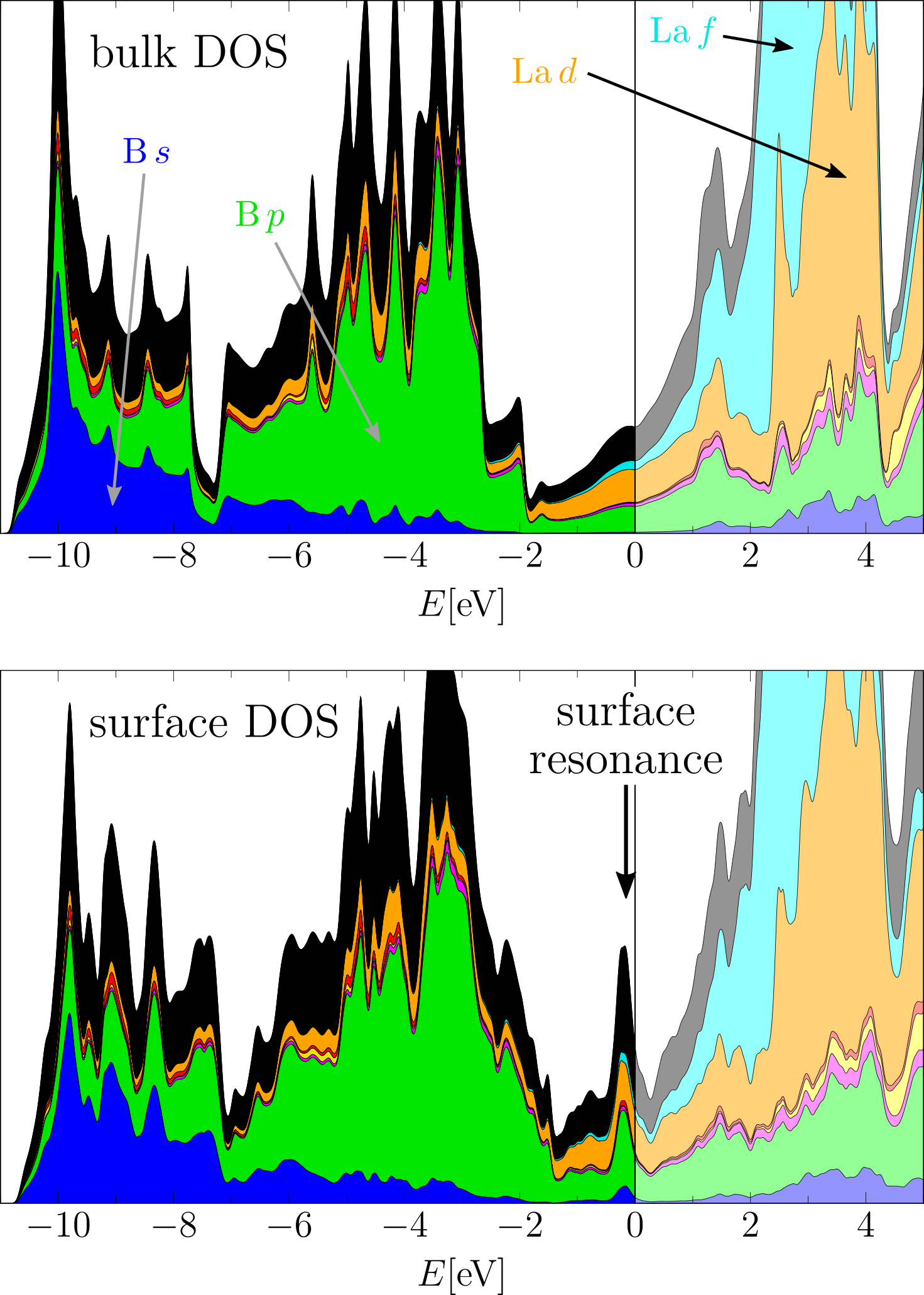}
\caption{Bulk DOS (top) and surface DOS (bottom) in arb.\ units projected onto atomic orbitals. The individual contributions are stacked. The Fermi energy is set to zero. Color code: total DOS (not stacked, black), B\,\(s\) (blue), B\,\(p\) (green), B\,\(d\) (magenta), La\,\(s\) (yellow), La\,\(p\) (red), La\,\(d\) (orange), La\,\(f\) (cyan). The characteristic surface peak is indicated by an arrow.}
\label{fig:dos_all}
\end{figure}

\begin{figure}[!hb]
\centering
\includegraphics[width=0.47\textwidth]{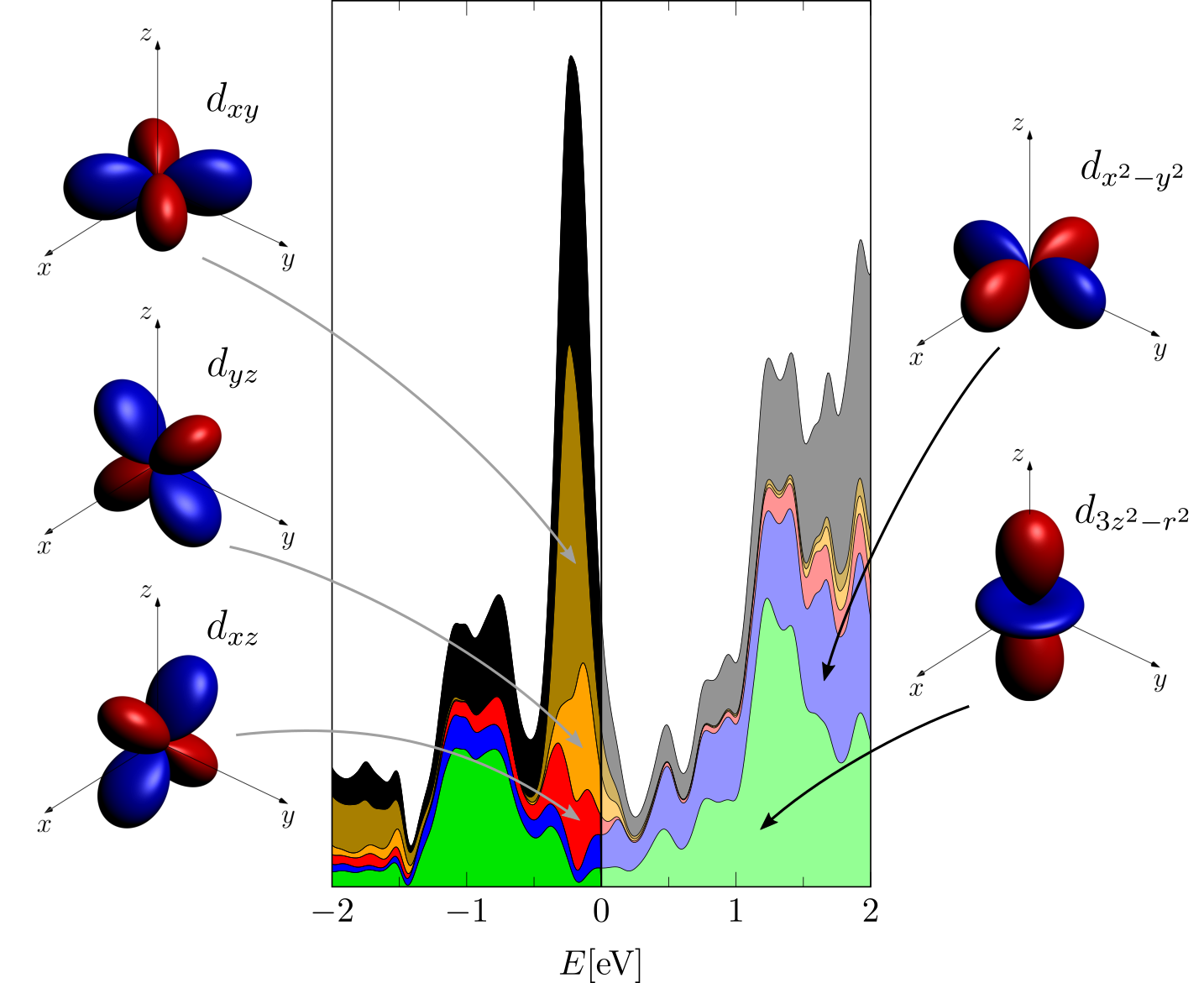}
\caption{DOS projected onto the \(d\)-orbitals of the La surface atoms, resolved by magnetic quantum number. DOS data in arb.\ units. The individual contributions are stacked. Color code: total DOS from La surface atoms (black), \(d_{xy}\) (gold), \(d_{yz}\) (orange), \(d_{3z^2-r^2}\) (green), \(d_{xz}\) (red), \(d_{x^2-y^2}\) (blue).}
\label{fig:dos_projection_d}
\end{figure}

\begin{figure*}[t]
\centering
\includegraphics[width=0.95\textwidth]{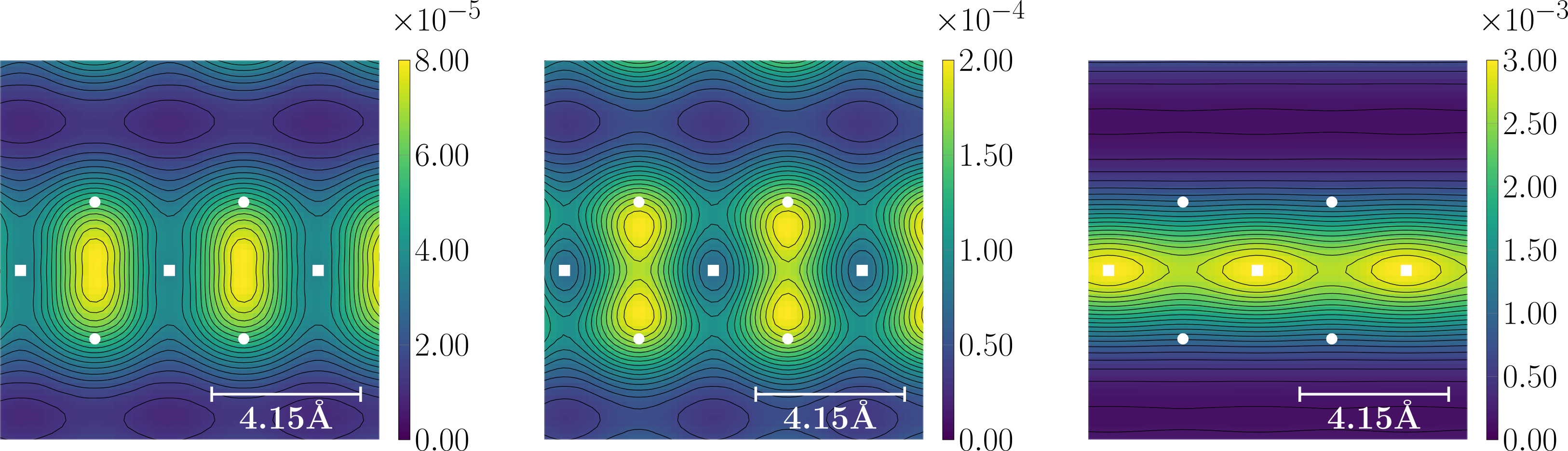}
\caption{Electron density at 4\,{\AA } above the surface for electrons from selected energy regions. Color scale in arb.\ units. Left: \(-1.29\)\,eV to \(-0.53\)\,eV, center: \(-0.53\)\,eV to \(0.28\)\,eV, right: \(0.28\)\,eV to \(1.64\)\,eV. The white squares mark the positions of lanthanum atoms at the surface. White circles indicate the positions of the topmost boron atoms.}
\label{fig:stm_contour_plots}
\end{figure*}

The angular momentum projected density of states (DOS) obtained from our DFT surface calculation is shown in \cref{fig:dos_all}. For comparison, we show the orbital projected DOS of a bulk simulation of LaB\(_6\) which is in good agreement with previous DFT results \cite{Uijttewaal, Ivashchenko}. In the data of the surface slab we find a characteristic peak \(-0.2\)\,eV below the Fermi level, which is not present in the bulk data. This peak is a surface feature and is composed largely of states of boron and lanthanum atoms closest to the surface. The peak is made of boron dangling bonds sticking out of the surface, which bind to the La \(d_{xy}\)-orbitals lying in the surface plane. The lobes of the \(d_{xy}\)-orbital point towards the four adjacent boron dangling bonds. While the contributions of the La \(d_{3z^2-r^2}\)- and La \(d_{x^2-y^2}\)-orbitals are rather small within the peak, they are dominant in the energy intervals adjacent to the peak. The projection of the DOS onto the \(d\)-orbitals of the La surface atoms is depicted in \cref{fig:dos_projection_d}. For additional DOS projections see \cref{sec:appendix_DFT_simulation_details}.

To connect our DFT simulations more closely to our STM/STS measurements, we follow Bardeen's tunneling theory \cite{bardeen} together with the arguments of Tersoff and Hamann \cite{tersoff1, tersoff2}, which relates the tunneling current for energies close to the Fermi level to the local density of states (LDOS) at the tip apex, integrated from the chemical potential of the probe to that of the tip.

Hence, in order to simulate the STM images, we compute the LDOS integrated over suitable energy windows. Rather than evaluating the LDOS at realistic tip positions, we choose a shorter distance of 4\,{\AA} above the plane of surface La ions. This is necessary, since at much larger distances the exponential decay of the LDOS leads to values, that are too small to be resolved in our calculations. Furthermore, at 4\,{\AA} the contrast of the significant features is particularly clear. However, by comparing the images taken at 4\,{\AA} with images calculated at larger distances, we ensured that the contrast does not change qualitatively. 

\cref{fig:stm_contour_plots} shows simulated STM images obtained from specific energy windows. At energies below \(E_\text{F}\), the simulation images are dominated by the boron lone pairs (\cref{fig:stm_contour_plots}, left and middle graph). The contrast is especially clear in the energy window covering the characteristic peak in the DOS 0.2\,eV below the Fermi level (\cref{fig:stm_contour_plots}, middle graph). While we observe distinct intensity maxima above the individual boron lone pairs in the energy window of the characteristic peak, we expect them to merge at larger distances, that correspond to realistic tip-surface distances.

The La ions are not visible in the energy windows below \(E_{\text{F}}+0.28\)\,eV, because the La \(d\)-orbitals are more localized than the boron lone pairs. In addition, the nodal structures of the La \(d\)-orbitals lead to a reduced density above the La ion in the range of the characteristic peak. This changes for the energy window above the characteristic peak at \(E_{\text{F}}-0.2\)\,eV: in the energy range from \(E_{\text{F}}+0.28\)\,eV to \(E_{\text{F}}+1.64\)\,eV, the La \(d_{3z^2-r^2}\)-orbitals pointing out of the surface contribute largely to the LDOS with the La \(f\)-orbitals providing an additional share. Hence the intensity is largest on top of the La positions.

In order to extract the local spectral information, which is provided by the scanning tunneling spectroscopy, we introduced so-called empty atoms above the surface. These atoms do not change the physics of the system, but they are a technical trick, that allows us to extract the LDOS in the vacuum region. The empty atoms provide local orbitals onto which the wave functions are projected to obtain the projected DOS.

\begin{figure}[b]
\centering
\includegraphics[width=0.4\textwidth]{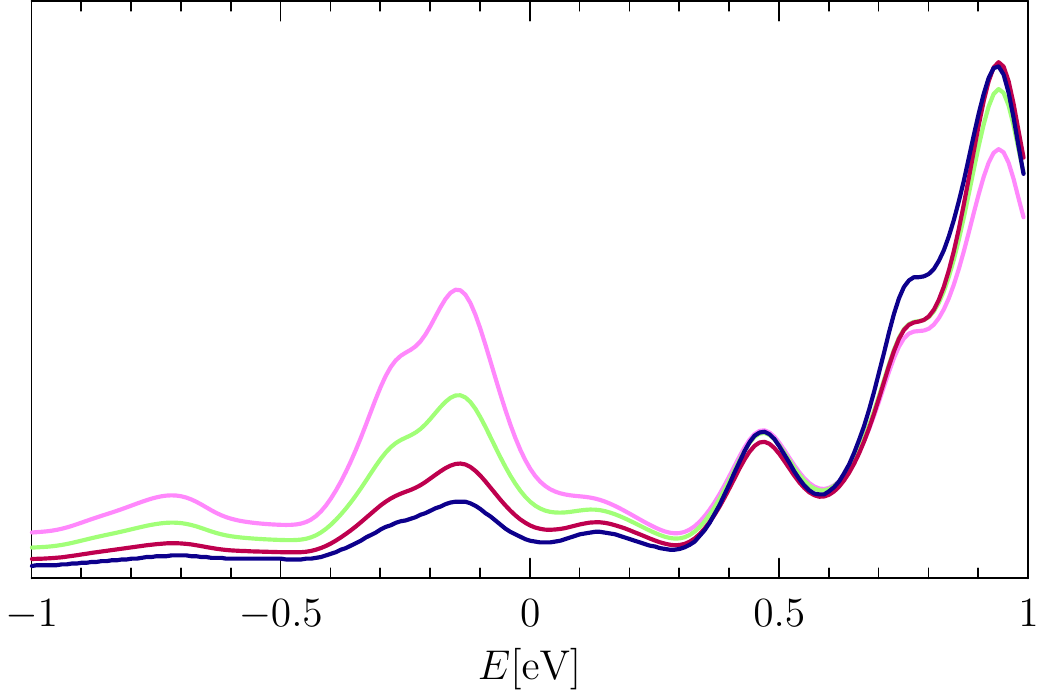}
\caption{DOS of the slab projected onto layers parallel to the slab at distances of 4\,{\AA} (pink), 5\,{\AA} (green, rescaled \(\times 6\)), 6\,{\AA} (red, rescaled \(\times 34\)) and 7\,{\AA} (blue, rescaled \(\times 200\)) above the surface. The graphs are in arb.\ units.}
\label{fig:dos_virtual_atoms}
\end{figure}

The LDOS projected onto layers of empty atoms is shown in \cref{fig:dos_virtual_atoms}. The characteristic peak of the boron dangling bonds below the Fermi level is very prominent. In addition we find a second peak at \(E_{\text{F}}+0.47\)\,eV. As seen in \cref{fig:dos_projection_d}, this peak is due to the \(d_{3z^2-r^2}\)-orbitals pointing out of the surface. At higher energies, in addition to the \(d_{3z^2-r^2}\)-orbitals, the \(f\)-orbitals start to contribute to the LDOS.

Above the peak at \(E_{\text{F}}+0.47\)\,eV, the LDOS rises sharply. The rise of the DOS at higher energies can be attributed to two reasons: Firstly, in this energy window the lanthanum ion has a large DOS due to \(d\)- and \(f\)-electrons. Secondly, wave functions at higher energies extend further out into the vacuum, because of their larger kinetic energy. 

By comparing the LDOS at different distances, we find that the intensity of the dangling bond peak decreases faster with increasing distance than the intensity of the LDOS above the Fermi energy. 
This can be seen in \cref{fig:dos_virtual_atoms}.

\section{Discussion}
By investigating the lanthanum hexaboride (001)-cleavage plane in UHV condition, we observed a chain-like (2\(\times\)1) reconstruction. 
Our STM experiments show that these chains are mostly labyrinth-like arranged and the number of parallel chains barely exceeds three or four. 
The most ordered (2\(\times\)1) reconstruction with a larger number of parallel chains can be found in the vicinity of step edges, where the chains tend to be ordered perpendicularly to the step edge. 
At a step edge which was rotated by 90$^{\circ}$, we observed the respective (1\(\times\)2) reconstruction. 
Following these arguments, the observed (1\(\times\)2) spots in the ULEED are likely caused by a step structure which was rotated by 90$^{\circ}$. 
This supports the assumption that the reconstruction aligns preferentially perpendicularly to step edges. 

The atomic surface structure of the (2\(\times\)1) reconstruction has been proposed as parallel lanthanum chains with a spacing of two lattice constants on top of a full B\(_6\) layer \cite{Schmidt}. Our DFT slab simulations of this lanthanum chain-terminated surface show that the electronic structure slightly below \(E_{\text{F}}\) is governed by a surface resonance. The orbital projected DOS reveals that the surface resonance is mainly composed of boron \(sp\)-hybrid dangling bonds pointing out of the surface. In contrast, in the unoccupied states, La \(d_{3z^2-r^2}\)-orbitals are predominant. In comparison to the boron surface states, these orbitals show a slower decay and extend further into the vacuum.

Our simulated STM images provide a translation of these results to the language of CCT images recorded by STM experiments. On the one hand, the dominant contribution of the La \(d_{3z^2-r^2}\)-orbitals to the LDOS at \(E>E_\text{F}\) leads to well separated chain-like structures above the surface. This is in good agreement with our experimental CCT images, which were taken at a positive bias voltages. Hence, at positive bias voltage, the STM addresses mainly the \(d_{3z^2-r^2}\)-orbitals. On the other hand, at \(E<E_\text{F}\) or negative sample bias voltage, respectively, our theoretical results predict tunneling predominantly from the boron lone pairs to the STM tip. This is difficult to realize experimentally, since we find rather unstable tunneling conditions for negative bias voltages. Hence, imaging the surface at negative bias voltages remains an open task.

The presence of a spectral feature below \(E_\text{F}\) is, however, verified by our STS measurements. In the tunneling spectra we find the surface resonance at \(-0.6\)\,eV and a parabola-shaped tunneling conductance around \(E_{\text{F}}\). Experimentally, an additional feature is found at 0.1\,eV. These features would coincide with the calculated DOS if \(E_{\text{F}}\) was shifted by \(+0.4\)\,eV in the calculated DOS. 
We tested the dependence of the surface resonance position on the density functional by doing additional calculations with the PBE functional, i.\,e. without Hartree-Fock contribution. 
The peak position does not change. 
This can be rationalized from the nature of the state, which is mostly boron-like and has little La $d$ and no La $f$ contribution. 

Although the investigated (2\(\times\)1) structure is far from pristine and exhibits numerous kinks and defects, one can exclude that the surface resonance position depends on the presence of surface defects. 
Experimentally we do not observe a shift of the surface resonance regardless of the local surface morphology, as seen in \cref{sts1V}. 
Furthermore, the main features of the experimental tunneling spectra can be clearly identified in our DFT results on the defect-free surface. 

A small shift in the chemical potential, which may explain the relative shift of 0.4\,eV between experimental and theoretical spectra, could originate from a non-stoichiometric crystal. During crystal growth, preferential evaporation of boron atoms leaves the formed crystal slightly boron deficient \cite{crystalgrowth}. Thus, the substance ratio La/B can be somewhat larger than the stoichiometric 1/6. For the present sample, the extra lanthanum atoms would lead to a surplus of charge carriers, shifting the Fermi level to higher energies. 

The measured apparent barrier height is rather small with values in the range from 1\,eV to 3\,eV. 
This is in agreement with the low work function of about 2.5\,eV obtained in former studies \cite{surfacetrenary}. 

Moreover, the knowledge that the (2\(\times\)1) reconstruction consists of lanthanum chains can be used to assign the termination of atomic steps. 
A step between two (2\(\times\)1) reconstructed areas should be of integer multiple height of the bulk lattice constant, since the termination is equal on each side. 
Such steps have been observed. 
For the area shown in the left of \cref{stepsB}, the step height is smaller than one bulk lattice constant. 
Here, we suggest a boron-rich termination with an almost vacant lanthanum layer at the surface. 
However, we would like to point out that this is one of the very few steps of that height we have found so far.

The surface morphology observed in this study is different to that of previous UHV STM investigations of LaB\(_6\) \cite{ozcomert1, ozcomert2}. 
In these studies, a (1\(\times\)1) reconstructed (001)-surface was found, which is stated to be lanthanum terminated and about 10\% of the surface's lanthanum sites are vacant. 
However, it should be noted that the samples used in these studies are prepared by polishing and heating instead of cleavage as in our study, which likely leads to the different surface structures observed. 
Our structural results are more comparable to the findings on the cleaved surface of SmB\(_6\). 
For cleaved SmB\(_6\), surface step heights are only of integer multiples of the bulk lattice constant if the terminations on both sides are of the same kind \cite{yee}, as observed in this study as well. 
Cleavage of SmB\(_6\) can take place by breaking either the bonds within the boron octahedra or the bonds between the octahedra. 
Hence, it is lacking a natural cleavage plane \cite{observation, ruan, matt}, and atomically ordered areas occur rather infrequently \cite{yee, jiao}. 
Since LaB\(_6\) has the same crystal structure, no natural cleavage plane is present. This might be the reason for the mostly disordered surfaces we have observed so far. 
Similarly to our observations, LEED investigations on cleaved SmB\(_6\) samples show only (2\(\times\)1) spots on certain surface areas and the diffraction pattern is governed by (1\(\times\)1) spots \cite{arpesstm}. 
Therefore, the (1\(\times\)1) spots have been associated with the bulk periodicity due to the finite penetration depth of the electrons. 
Same arguments should be applied for our sample system to explain the dominant (1\(\times\)1) spots.

\section{Conclusion}
In a combined study of STM/STS, LEED and DFT we have investigated the (001)-cleavage plane of LaB\(_6\). 
Atomically ordered areas are labyrinth-like (2\(\times\)1) reconstructed. 
These chains can be understood as parallel rows of lanthanum atoms on top of a B\(_6\) layer, with a surface resonance below \(E_{\text{F}}\). 
Electronically, this resonance is mainly composed of B \(sp\)-hybrid orbitals and La \(d_{xy}\)-states originating from the surface atoms. 
Lanthanum hexaboride is the electronically most simple candidate of the REB\(_6\) family. 
However, understanding the electronic surface structure is not straightforward and differs severely from that of the bulk. 
Moreover, the LaB\(_6\) surface morphology is more complex than previously discussed. 
Since all REB\(_6\) have the same crystal structure, our findings could help to further understand the general surface physics of hexaborides. 

\begin{acknowledgments}
We acknowledge fruitful discussions with S.\ R.\ Manmana. The orientation of the single crystals via gamma ray diffraction were carried out in the group of G.\ Eckold by F.\ Ziegler and P.\ Kirscht at the Institut für Physikalische Chemie, University of Göttingen. We gratefully acknowledge financial support by the DFG grants WE1889/10-1, BL539/10-1 and PR298/19-1. Furthermore, we acknowledge financial support by the SFB 1073 through the projects B03, C03 and C04. Additionally, this work was funded by the European Research Council (ERC Starting Grant ‘ULEED’, ID: 639119). The work at the University of Warwick was supported by the EPSRC, UK, through Grant EP/M028771/1. The images in \cref{schema}, \cref{fig:unitcell_x} and \cref{fig:structure_relaxation} were created with VESTA \cite{momma}. 
\end{acknowledgments}

~\\
F. Sohn and P. Buchsteiner contributed equally to this work.

\begin{appendix}
\section*{appendix}
\subsection{Supporting experimental data}
\label{Supporting_experimental_data}
\begin{figure}[ht]
\centering
\includegraphics[scale=0.16]{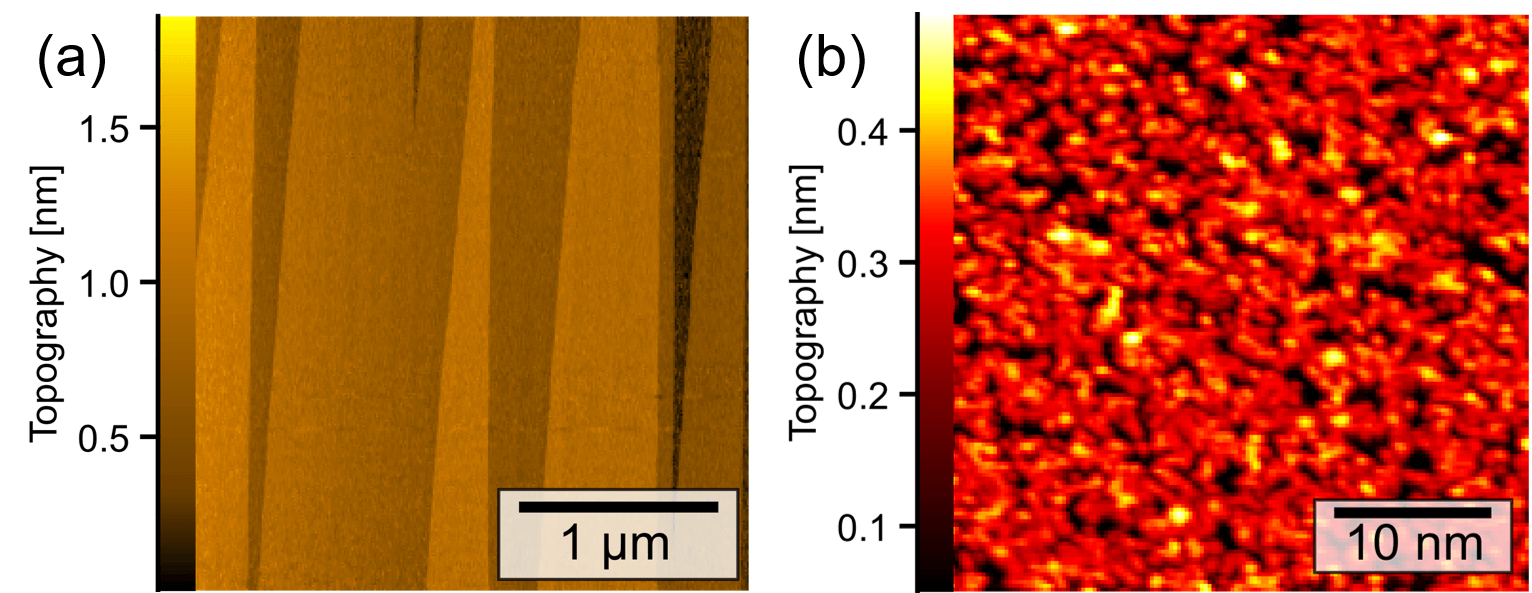}
\caption{(a) (001)-surface of cleaved LaB\(_6\) as observed with AFM under ambient conditions. (b) CCT of a disordered surface area taken at 1\,V/0.1\,nA.}
\label{surfacemorphologies}
\end{figure}

\cref{surfacemorphologies}\,(a) shows a typical large scale topography of the cleaved LaB\(_6\) (001)-surface as observed with AFM. 
Flat terraces within the AFM resolution with areas of several hundred (nm)\(^2\) up to (\(\mu\)m)\(^2\) in size can be easily found. 
These areas are connected by steps of integer multiples of the bulk lattice constant of 4.15\,{\AA}  \cite{stackelberg, guo, booth, sirota, gurel, lattice, Schmidt}. 
The steps in \cref{surfacemorphologies}\,(a) are all of a height of one lattice constant. 
However, on a smaller length scale probed by STM, up to nanometer-sized protrusions with no pristine long-range order are commonly observed, \cref{surfacemorphologies}\,(b). 
This results are similar to the STM findings on cleaved SmB\(_6\) \cite{yee}. 

As described in the discussion above, LaB\(_6\) is lacking a natural cleavage plane. 
If the cleavage leaves the boron cage intact, one dangling bond per octahedron is exposed. 
The situation changes dramatically, if the cleavage proceeds through the octahedron. 
The disrupted boron cages expose numerous dangling bonds, which might increase the probability of attaching adsorbates like hydrogen \cite{hydrogen}. 
Another reason for a frequently observed disordered morphology might be that the crystals were cleaved at room temperature. 
For SmB\(_6\) it is reported, that atomically flat surfaces could not be obtained by cleavage at room temperature, but only at about 20\,K \cite{dresden}. 
Our present setup does not allow the cleavage at cryogenic temperature. 
An additional experimental reason for the widely disordered appearance might be the tip itself. 
As seen in \cref{FFT}, even when atomic chains can be resolved, they are mostly labyrinth-like arranged. 
If the same area would have been scanned with a slightly blunt tip, the topography certainly would appear rather disordered, too. 
\begin{figure}[h]
\centering
\includegraphics[width=0.43\textwidth]{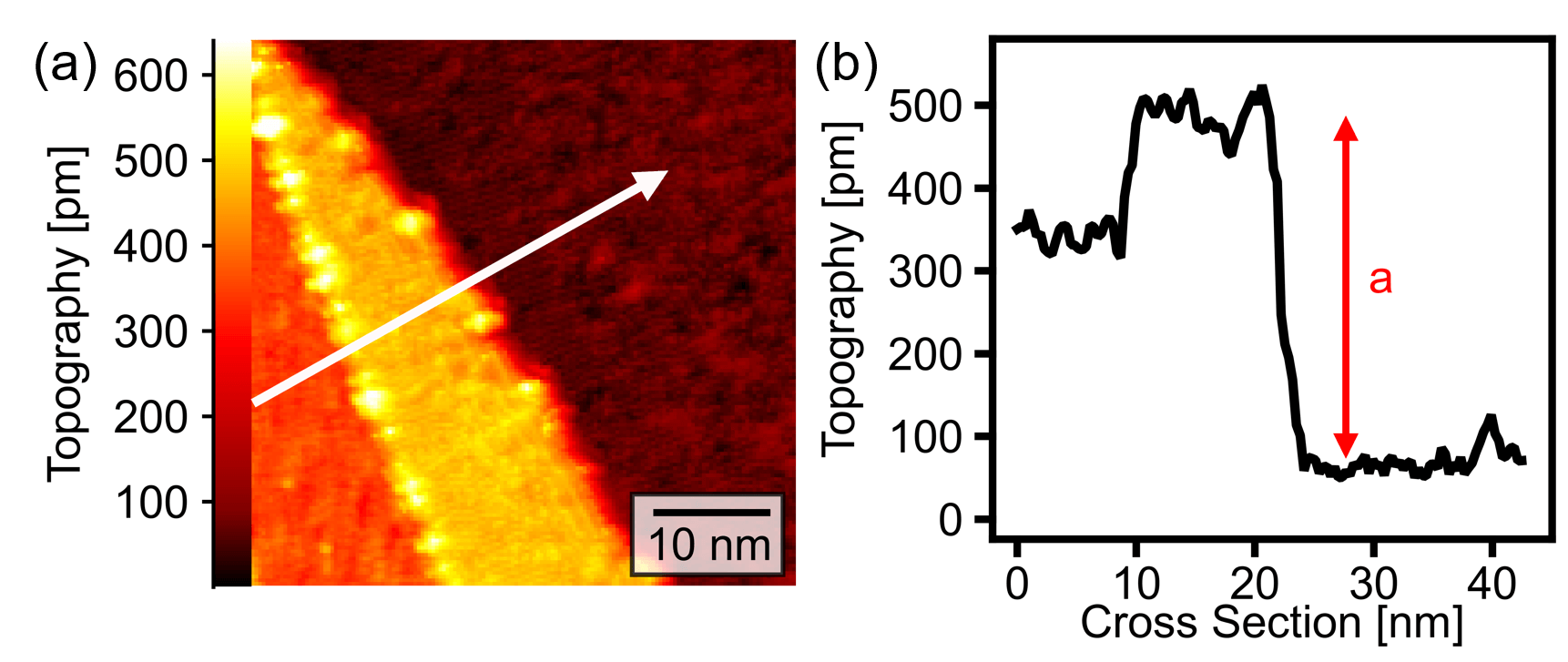}
\caption{(a) CCT image taken at 0.8\,V/0.1\,nA of two steps, one of about 1.3\,\AA\;height and one of a bulk lattice constant height, as shown in the height profile in (b).}
\label{stepsB}
\end{figure}

\begin{figure}[b]
\centering
\includegraphics[scale=0.135]{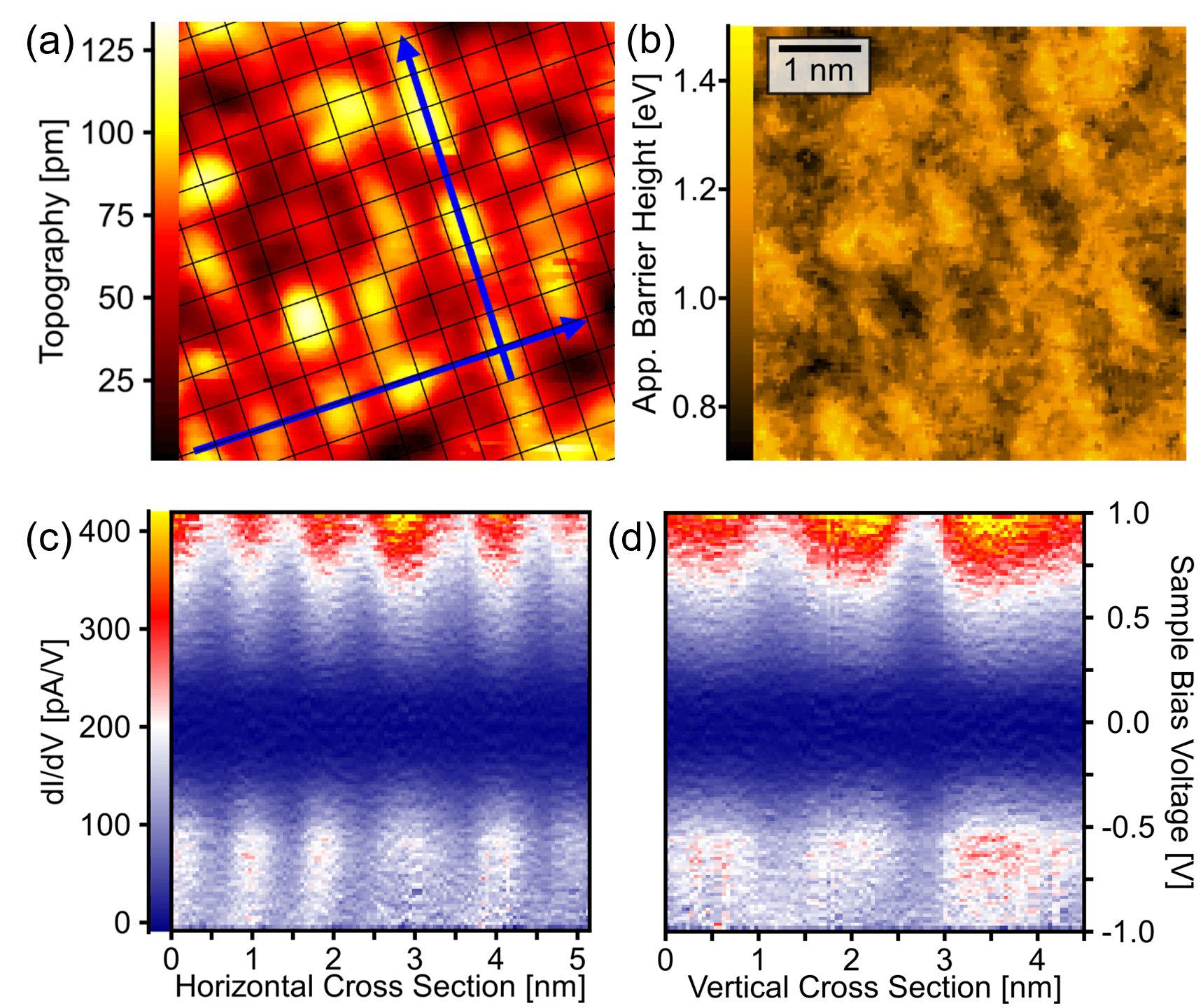}
\caption{(a) The CCT of \cref{dIdV} is shown again, this time with an overlay of the atomic lattice grid for scale. The blue arrows mark the lines along which the horizontal and vertical cross sections are taken. The resulting color coded d\(I/\)d\(V(V,x)\) maps can be seen in (c) and (d). In (b) the spatially resolved apparent barrier height is depicted.}
\label{sts1V}
\end{figure}

We have found a step with a height of only a fraction of the bulk lattice constant, which is shown in \cref{stepsB}. 
On the right hand side the already discussed (2\(\times\)1) labyrinth-like structure is observed and the step height is again one bulk lattice constant. 
Interestingly, the step on the left hand side is only about 1.3\,\AA\;high. 
We interpret this finding as a different surface termination, as described in the discussion above. 

\cref{sts1V} shows the spectroscopy of \cref{dIdV} again, but here including the measured apparent barrier height and another d\(I/\)d\(V(V,x)\) cross section. 
The mean value of the \(\Phi_{\text{app}}\)-map is (1.05 \(\pm\) 0.17)\,eV. 
However, the specific value of the apparent barrier height is correlated with the topography. 
At the atomic protrusion, blue-marked position in \cref{dIdV}\,(a), its value is about 1.1\,eV and at the green-marked position 1.0\,eV. 
For the various surface areas with no clear atomic rows included, the value of \(\Phi_{\text{app}} (x,y)\) varies for each tip position. 

\begin{figure}[ht]
\centering
\includegraphics[scale=0.13]{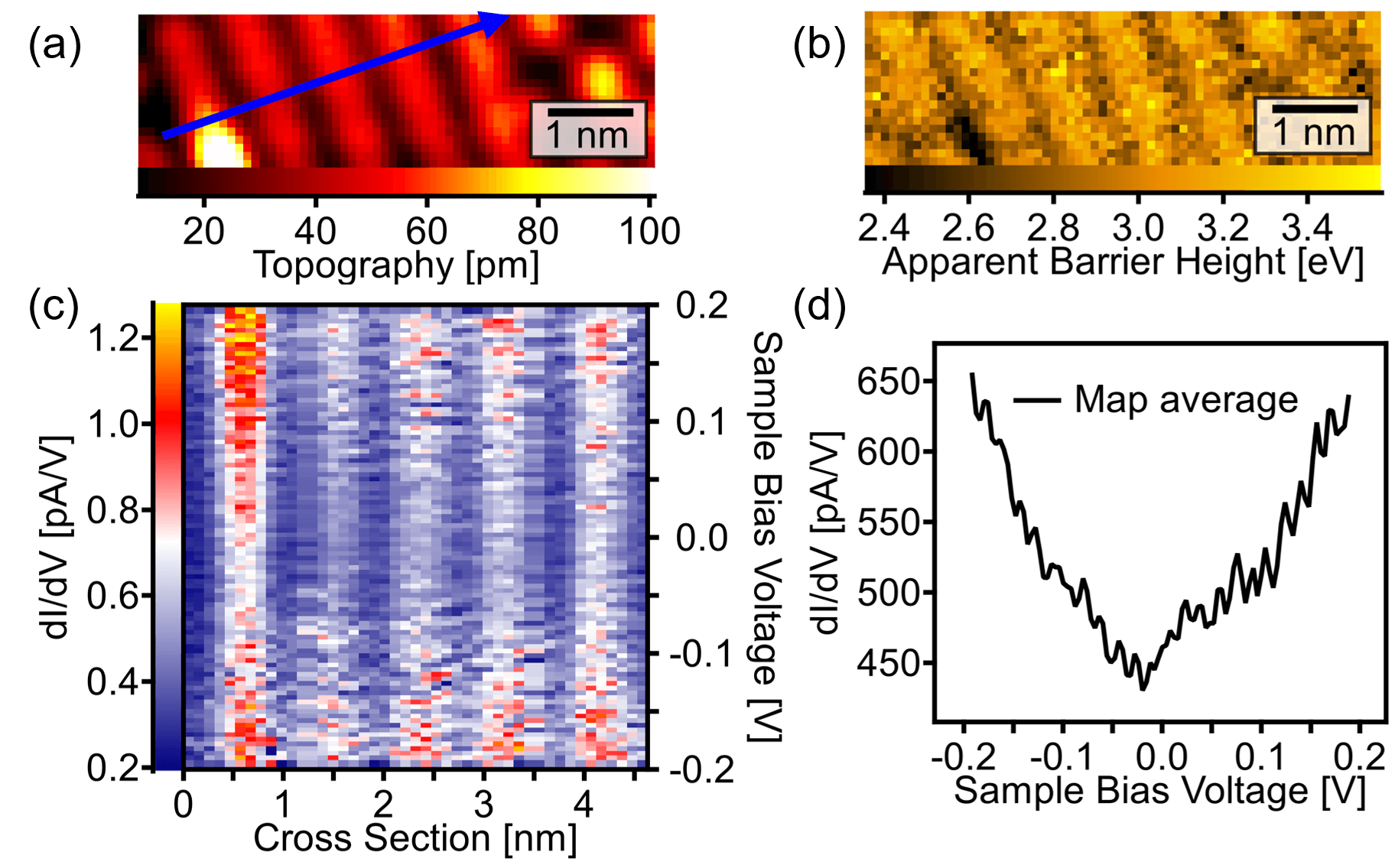}
\caption{(a) CCT taken at 0.2\,V/0.1\,nA. The apparent barrier height is shown in (b). The color coded d\(I/\)d\(V(V,x)\) cross section taken along the blue arrow within (a) is depicted in (c). (d) shows the map averaged d\(I/\)d\(V(V)\) curve of (a).}
\label{sts200mV}
\end{figure}

\cref{sts200mV} shows the second spectroscopy data set, including the CCT, the \(\Phi_{\text{app}} (x,y)\)-map and a d\(I/\)d\(V(V,x)\) cross section after topography normalization. 
In this bias voltage range, the d\(I/\)d\(V(V)\) curves have their largest values on top of the reconstruction's protrusion.

\subsection{Additional detail on the DFT simulations}
\label{sec:appendix_DFT_simulation_details}
\begin{figure}[h]
\centering
\includegraphics[width=.34\textwidth]{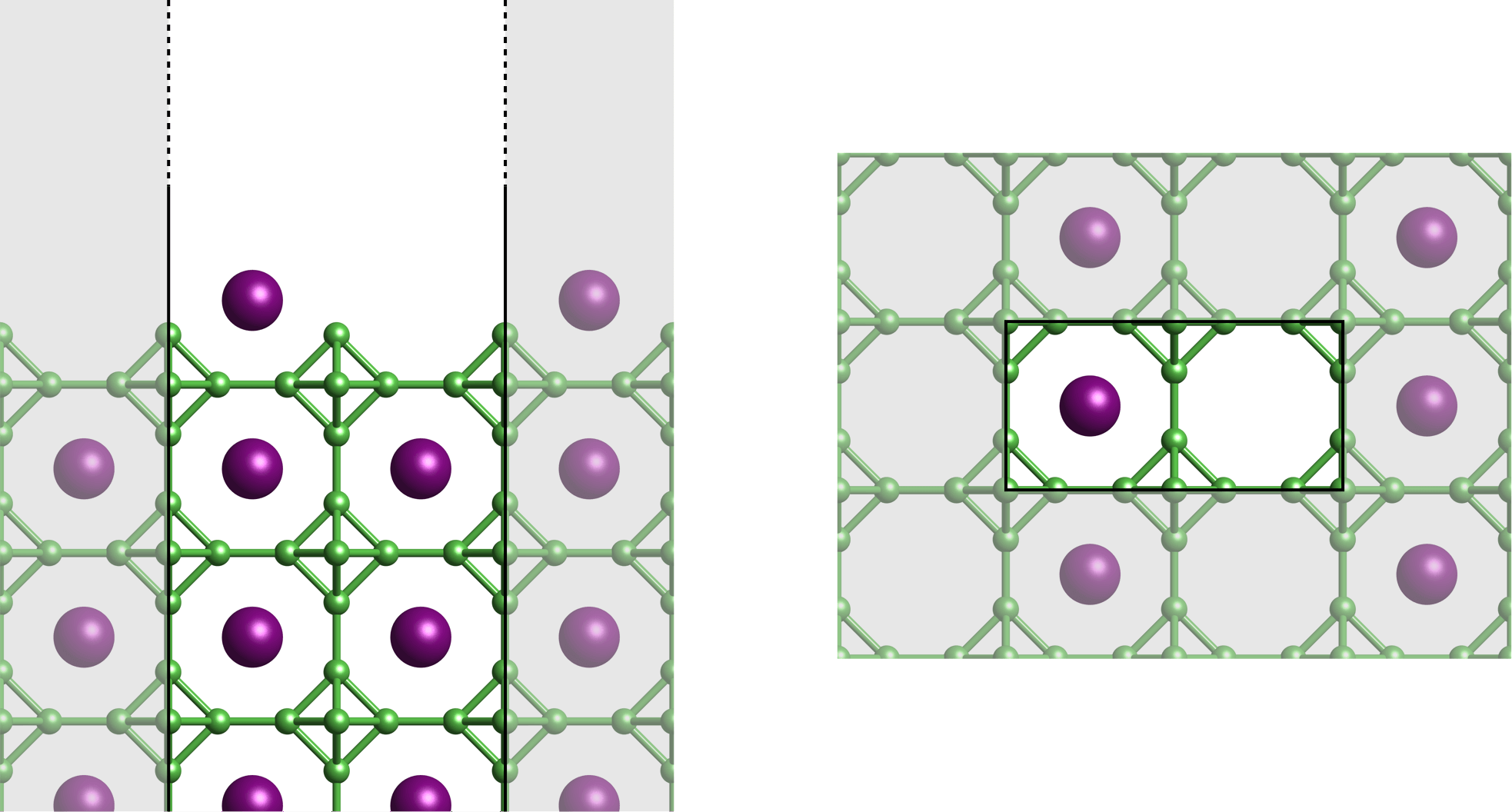}
\caption{Side and top view of the (001)-surface of LaB\(_6\). The slab unit cell is highlighted by a white background and marked with the black lines. On the surface only every second La row is filled.}
\label{fig:unitcell_x}
\end{figure}

A scheme of a simulated slab and its (2\(\times\)1) surface unit cell is depicted in \cref{fig:unitcell_x}. 

The lattice constant was kept fixed at a value of \(a=4.15\)\,{\AA} found in former experimental and theoretical studies \cite{stackelberg, guo, booth, sirota, gurel, lattice, Schmidt}. The vacuum thickness between the slabs was set to \(16.6\)\,\AA. In order to avoid artifacts and to have a sufficiently large spatial separation of the slab's top and bottom surfaces and surface states, respectively, a minimum slab thickness of three layers of boron octahedra is required. For our simulations, we used surface supercells with three, five and seven layers of boron octahedra. This was done in order to ensure convergence of our DFT data with respect to the slab thickness. In this work, we present results from our simulations with the seven-layer slab unit cell only. However, all slab thicknesses lead to similar results in atomic structure as well as electronic structure.

Moreover, we used a total amount of 8\(\times\)4\( \times\)1 \(\mathbf{k}\)-points. The plane wave cut-offs were set to 50\,Ry for the wave functions and to 100\,Ry for the charge density. Furthermore, we used (2,2,2,1) projector functions for the La (\(s\), \(p\), \(d\), \(f\))-states and (2,2,1) projector functions for the B (\(s\), \(p\), \(d\))-states, respectively. The hybrid functional mixing factor was fixed at 0.15 for both atomic species. 

\begin{figure}[h]
\centering
\includegraphics[width=0.29\textwidth]{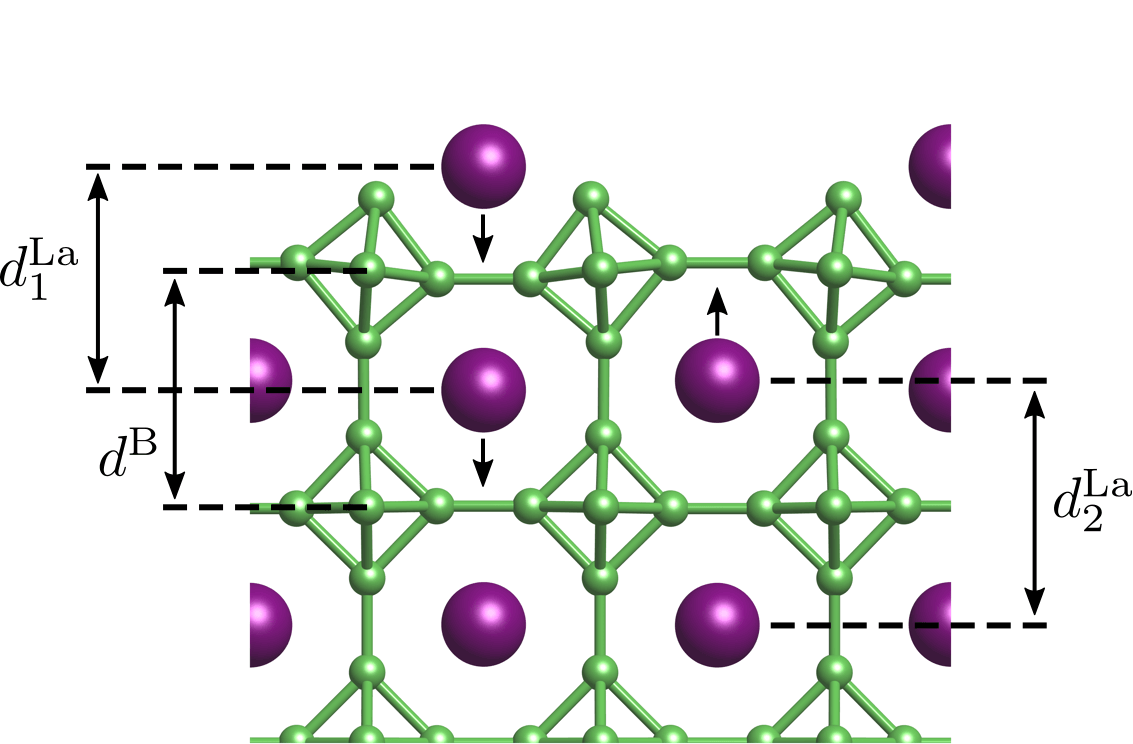}
\caption{Structure relaxation at the slab surface. \(d^\text{La}\) are distances between La atoms of the topmost layers as indicated. \(d^\text{B}\) is the distance between the centers of mass of the two octahedra. The small black arrows indicate the direction of the relaxation movement compared to the bulk structure.}
\label{fig:structure_relaxation}
\end{figure}

In \cref{fig:structure_relaxation} we show the side view of the relaxed surface structure, which exhibits a tilt of the surface octahedra towards the filled chain of La ions. This can be rationalized by the attraction between positive La and negative B ions: the terminating boron atoms have lone-pairs pointing out of the surface, which hybridize with the La \(d_{xy}\)-orbitals lying in the surface plane (cf. \cref{fig:dos_projection_d} and discussion in \cref{sec:surface_simulations}).

Furthermore, the surface relaxation leads to changes in the distance between two layers which can be quantified via the relative change to the bulk lattice constant \(a_\text{bulk}\), i.\,e.,
\begin{equation}
\Delta = \frac{d-a_\text{bulk}}{a_\text{bulk}}.
\end{equation}
The surface La-ion relaxes inward by \(\Delta^{\text{La}}_1=-4.3\)\,\% relative to the bulk, while the subsurface La-ion below the empty La-surface site relaxes outward by \(\Delta^{\text{La}}_2=+4.7\)\,\%. The boron octahedra exhibit a small outward relaxation of \(\Delta^{\text{B}}=+1.4\)\,\%. A comparison of these quantities between the data from the seven-B\(_6\)-layer slab to the data of the five-B\(_6\)-layer slab shows only minor differences which amount up to 0.007\,{\AA} in absolute values, which is negligible in the context of our calculations. Hence, we expect no significant difference in the DFT results for even larger slab thicknesses. Previous data published by \citet{Schmidt} are \(\Delta^{\text{La}}_1 = -6.2\)\,\%, \(\Delta^{\text{La}}_2 = +3.5\)\,\%, and \(\Delta^{\text{B}} = +0.4\)\,\%, respectively. In comparison, the distances in our data are consistently a bit longer.

\begin{figure}[h]
\centering
\includegraphics[width=0.38\textwidth]{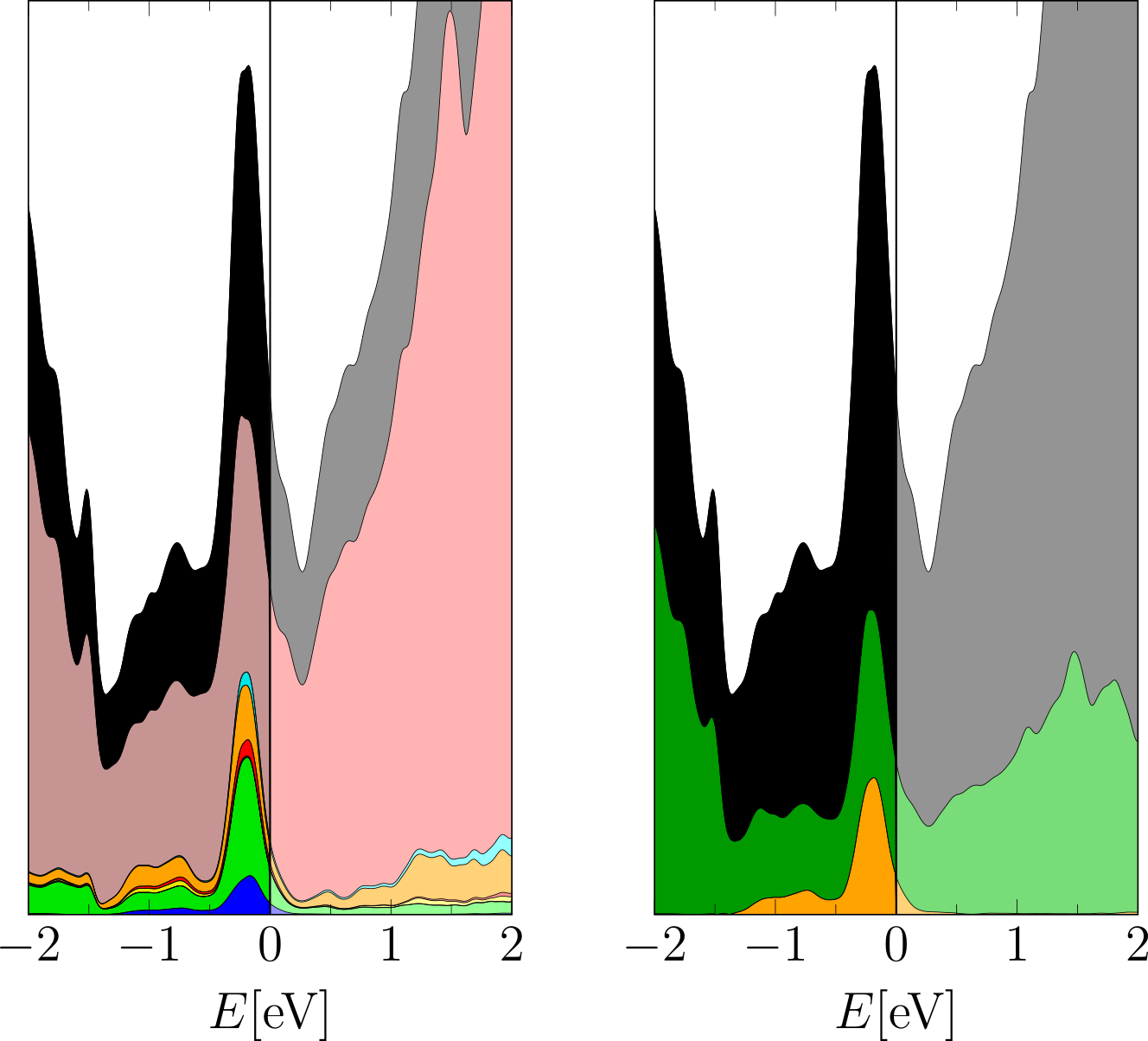}
\caption{Projections of the DOS onto orbitals of the surface layer atoms. DOS data in arb.\ units. The individual contributions are stacked. Left: DOS projected onto the topmost boron and lanthanum atoms. The color code is identical to the color code in \cref{fig:dos_all}. The bulk contribution is stacked on top (reddish brown). Right: DOS projected onto boron orbitals. The dangling bonds of the boron surface atoms (orange) dominate the peak close to the Fermi level. The contribution of all other boron orbitals of the slab (dark green) is shown for comparison.}
\label{fig:dos_projection_surface_all_and_surface_boron}
\end{figure}

In \cref{fig:dos_projection_surface_all_and_surface_boron}, we provide two additional projections of the DOS: a projection onto the orbitals of the topmost atoms of either species, and a projection onto the boron dangling bonds.

\end{appendix}

\bibliographystyle{apsrev4-2}
\bibliography{references}

\end{document}